\documentclass[pra,showpacs,showkeys,amsfonts,amsmath,twocolumn]{revtex4}
\usepackage{bm}
\usepackage{graphicx}
\RequirePackage{mathptm}

\numberwithin{equation}{section}
\newcommand{\si}[1]{\sigma_{#1}}
\newcommand{\sib}{\boldsymbol{\sigma}}

\newcommand{\ip}[2]{\langle \,{#1},\,{#2}\,\rangle}

\newcommand{\ro}{\varrho}
\newcommand{\rocq}{\varrho_{\mathrm{cq}}}
\newcommand{\rocc}{\varrho_{\mathrm{cc}}}

\newcommand{\la}[1]{\boldsymbol{\lambda}_{#1}}

\newcommand{\al}[1]{\alpha_{#1}}

\newcommand{\om}{\omega}

\newcommand{\te}{\vartheta}
\newcommand{\vf}{\varphi}

\newcommand{\I}{\openone}

\newcommand{\C}{\mathbb C}
\newcommand{\R}{\mathbb R}

\newcommand{\PP}{\mathbb P}

\newcommand{\cA}{{\mathcal A}}
\newcommand{\cB}{{\mathcal B}}
\newcommand{\Atot}{{\mathcal A}_{\mathrm {tot}}}
\newcommand{\tr}{\mathrm{tr}\,}

\newcommand{\tl}[1]{\boldsymbol #1}
\newcommand{\mr}[1]{\mathrm{#1}}

\newcommand{\sr}[1]{\langle  {#1}\rangle}
\newcommand{\DS}{\displaystyle}

\newcommand{\cm}{C_{\mathrm{M}}}
\begin{document}
\title{Maximal mutual correlation: an algebraic measure of correlations in bipartite quantum systems}
\author{  Lech Jak{\'o}bczyk
\footnote{ ljak@ift.uni.wroc.pl} }
 \affiliation{Institute of Theoretical Physics\\ University of
Wroc{\l}aw\\
Plac Maxa Borna 9, 50-204 Wroc{\l}aw, Poland}
\begin{abstract}
We introduce an algebraic measure of correlations in bipartite
quantum systems. The proposed quantity, called maximal mutual
correlation, provides  the information how much a given state
differs from the product state of its marginals. In contrast to the
entanglement or quantum discord, maximal mutual correlation can be
non - zero for some classically correlated quantum states.
\end{abstract}
\pacs{03.67.Mn,03.65.Yz,42.50.-p} \keywords{bipartite quantum
systems, algebraic measure of correlations, quantum discord}
\maketitle
\section{Introduction}
Various notions of quantumness of correlations in a given quantum state give essentially the
same information in the case of pure states. A pure state of bipartite quantum system is either
separable (i.e. a product state) or nonseparable. This mathematical notion of nonseparability can be extended to
the mixed states, giving the natural (but formal) definition of entanglement of general quantum states:
a (mixed) state is entangled if it cannot be expressed as a convex combination of pure separable states \cite{W}.
It was widely accepted that such defined entanglement is the only source of quantumness of correlations. In the recent
years however, other, more general  features of correlations have attracted much interest. They arise from the observation that
for pure separable state, there exists  von Neumann measurement on a part of composite system that do not disturb the state, whereas nonseparable states are always disturbed by such local  measurements . Extension of this feature to the mixed states, gives rise of the notion of quantum discord \cite{Z,HV,M}.
For pure states discord coincides with entanglement, but in the case of mixed states discord and entanglement differ significantly.
For example,  almost all quantum states have non-vanishing discord and in particular there exist discordant separable
mixed states \cite{F}.
\par
A pure separable state of bipartite systems has also  important
algebraic property: the expectation value $\om$ in such state
factorizes i.e.
\begin{equation}
\om(AB)=\om(A)\om(B)\label{prod}
\end{equation}
for the product of all  local observables. On the other hand, for
nonseparable states there are such pairs of local observables that
(\ref{prod}) is not valid. Therefore the maximum of the numbers
\begin{equation}
|\om(AB)-\om(A)\om(B)|\label{omab}
\end{equation}
taken over all pairs of local observales, can serve as a kind of an
algebraic measure of quantumness of correlations. This idea was
applied to the case of two qubits in Ref. \cite{DGJ}, where it was
shown that the suitably normalized maximum of (\ref{omab}) exactly
equals to the pure state entanglement. In the present note we study the extension of
that measure of correlations to the mixed two-qubit states. The extended measure we call
\textit{maximal mutual correlation} between subsystems in a given state. In contrast to the entanglement
or quantum discord, this quantity can be non - zero even for classically correlated (i.e. non - discordant)
quantum states.
\par
In this article we compute maximal mutual correlation for several
classes of states and study the relation of this quantity to other
measures of correlations in two - qubit quantum systems. In
particular we show that in the case of so called "classical -
quantum" states, this measure depends on the classical probability
distribution defining such states but also encodes some information
about quantum states of the second subsystem, whereas for "classical
- classical" states it depends only on the properties of the
corresponding classical probability distribution. In the  states
containing some quantum correlations, maximal mutual correlation is
always greater or equal to geometric measure of quantum discord.
This relation is proved to be valid in the class of Bell - diagonal
states
 and it is conjectured that  is also true for general two - qubit states.

\section{Bipartite correlations in the algebraic framework}
\subsection{General discussion}
For the unambiguous discussion of quantum non-separability, the crucial is a clear
identification of subsystems of the total system. In the context of algebraic quantum
theory such identification is given  by the proper choice of the factorization of the total
algebra  of observables into subalgebras describing subsystems (see e.g. \cite{B}).
In the algebraic formulation of quantum theory (see e.g. \cite{Emch}), the total system is described by the
$\ast$ - algebra $\Atot$ of all observables. Since in the present note we consider only
the systems with finite number of levels, $\Atot$ is isomorphic to the full matrix algebra.
The most general state on $\Atot$ is given
by the linear functional
\begin{equation}
\om\,:\,\Atot\to \C,\label{state}
\end{equation}
which is positive i.e. for all $A\in \Atot$
\begin{equation*}
\om(A^{\ast}A)\geq 0,
\end{equation*}
and normalized  i.e.
\begin{equation*}
\om(\I)=1.
\end{equation*}
For the matrix algebras, any state $\om$ is normal i.e. there exists density matrix $\ro$ such that
\begin{equation}
\om(A)=\tr (\ro A)\label{mstate}
\end{equation}
To describe subsystems of the total system, we consider subalgebras
$\cA$ and $\cB$ of $\Atot$. We assume  that $\cA\cap \cB=\{c\I\}$
and  $\Atot=\cA\vee \cB$ i.e. $\Atot$ is generated by $\cA$ and
$\cB$. For the discussion of  correlations between observables in
$\cA$ and $\cB$ in a given state $\om$, the crucial is the notion of
uncorrelated states. The formal definition is as follows: the state
$\om$ on $\cA\vee \cB$ is $(\cA,\, \cB)$ - \textit{uncorrelated} if
\begin{equation}
\om(AB)=\om(A)\,\om(B)
\end{equation}
for every $A\in \cA$ and $B\in \cB$. As it was shown in Ref.
\cite{BS} the very existence of such uncorrelated or product states
depends on the mutual commutativity of the algebras $\cA$ and $\cB$.
Thus  in the context of the theory  of  correlations in bipartite
systems, the total algebra of observables should be such that
 $\Atot= \cA\vee \cB$, where $ \cA\cap \cB=\{c\I\}$ and for all $A\in \cA$ and $B\in \cB,\; [A,B]=0$.
In the case of full matrix algebras, much more can be said about the structure of $\Atot$. One can show that
$\cA\vee\cB$ is isomorphic to the tensor product $\cA\otimes \cB$. So  $\cA$ is generated by elements $A\otimes \I$
and  $\cB$ is generated by $\I\otimes B$.
\par
Now we pass to the discussion of correlations between subsystems. A
state $\om$ on $\Atot$ is $(\cA,\, \cB)$ - \textit{correlated} if it
is not a product state with respect to subalgebras $\cA$ and $\cB$.
Next we introduce a quantity which we call maximal mutual correlation
between $\cA$ and $\cB$ in $\om$. It is defined as \cite{DGJ}
\begin{equation}
\cm(\om)=\sup\limits_{A,B}\,|\om(AB)-\om(A)\om(B)|.\label{MMC}
\end{equation}
In the formula (\ref{MMC}), the supremum is taken over all normalized elements $A\in \cA$ and $B\in \cB$.
The quantity $\cm(\om)$ gives the information how much a state differs from the product state of
its marginals, when we take into account only local measurements. Since the states $\om$ are of the form (\ref{mstate}),
we put $\cm(\om)=\cm(\ro)$. Notice that
\begin{equation}
\cm(\ro)\leq ||\om-\om^{\cA}\otimes \om^{\cB}||=||\ro-\ro^{\cA}\otimes\ro^{\cB}||_{1},
\end{equation}
where $||\cdot||_{1}$ is a trace norm and the marginal states $\om^{\cA}$ and $\om^{\cB}$ are given by
partial traces $\ro^{\cA}$ and $\ro^{\cB}$ of a density matrix $\ro$.
Thus $\cm(\ro)$ is always dominated by so called \textit{correlation distance} $C(\ro)$, given by
\begin{equation}
C(\ro)=||\ro-\ro^{\cA}\otimes\ro^{\cB}||_{1} \label{cd}
\end{equation}
and introduced recently by Hall \cite{Hall} in the context of the analysis of quantum mutual information.
\subsection{Two qubit case}
In the case of two qubits, the total algebra $\Atot$
can be considered as generated by matrix unit $\I$ and elements
$\la{1},\ldots,\la{15}$, where
\begin{equation}
\la{i}=\I\otimes \si{i},\quad  \la{3+i}=\si{i}\otimes \I,\quad
i=1,2,3
\end{equation}
and $\la{j},\; j=7,\ldots,15$ are given by tensor products of the
Pauli matrices $\si{i}$ taken in the lexicographical order. So
\begin{equation}
\Atot=\left[\,\I,\la{1},\ldots,\la{15}\,\right],
\end{equation}
and
\begin{equation}
\cA=\left[\, \I,\, A_{1},\, A_{2},\, A_{3}\,\right],\quad
\cB=\left[\, \I,\, B_{1},\, B_{2},\,
B_{3}\,\right],\label{A0B0}
\end{equation}
where
\begin{equation}
A_{i}=\la{3+i},\; B_{i}=\la{i},\quad i=1,2,3.
\end{equation}
It is convenient to take elements $A\in \cA$ and $B\in \cB$ defined as
\begin{equation}
A=a_{1}A_{1}+a_{2}A_{2}+a_{3}A_{3},\quad B=b_{1}B_{1}+b_{2}B_{2}+b_{3}B_{3},
\end{equation}
where $\tl{a}=(a_{1},a_{2},a_{3}),\; \tl{b}=(b_{1},b_{2},b_{3})$
are normalized vectors in $\R^{3}$. Then $A^{2}=\I,\; B^{2}=\I$ and
\begin{equation}
\om(AB)-\om(A)\om(B)=\ip{\tl{a}}{Q\,\tl{b}},
\end{equation}
where the covariance matrix $Q=(q_{ij})$ has the matrix elements
\begin{equation}
q_{ij}=\tr(\ro A_{i}B_{j})-\tr(\ro A_{i})\tr(\ro B_{j}).
\end{equation}
So \cite{DGJ}
\begin{equation}
\cm(\ro)=\sup\limits_{||\tl{a}||=||\tl{b}||=1}|\ip{\tl{a}}{Q\,\tl{b}}|.\label{cm2q}
\end{equation}
Notice that the right hand side of (\ref{cm2q}) is the norm of the matrix $Q$, thus for two qubits
\begin{equation}
\cm(\ro)=\max\,(t_{1},\, t_{2},\, t_{3}),
\end{equation}
where $\{t_{1},\, t_{2},\, t_{3}\}$ are singular values of the covariance matrix $Q$.
\par
In the two - qubit case, there is also an explicit expression for the correlation distance $C(\ro)$ \cite{Hall}.
 Again it is given in terms of singular values of $Q$
\begin{equation}
\begin{split}
C(\ro)=&\frac{1}{4}\big(|t_{1}+t_{2}+t_{3}|+|t_{1}+t_{2}-t_{3}|+|t_{1}-t_{2}+t_{3}|\\
&\hspace*{4mm}+|-t_{1}+t_{2}+t_{3}|\big)
\end{split}
\end{equation}
\section{Relation of $\cm(\ro)$ to other measures of correlations}
In the general case, the maximal mutual correlation $\cm(\ro)$ is
upper bounded by the correlation distance $C(\ro)$. In this section
we focus on the case of two qubits and compare $\cm(\ro)$ with other
measures of correlations corresponding to entanglement and quantum
discord. As a measure of entanglement we choose normalized
negativity $N(\ro)$, given by
\begin{equation}
N(\ro)=||\ro^{PT}||_{1}-1,
\end{equation}
where $PT$ stands for partial transposition. To quantify quantum
discord, we take geometric measure of quantum discord defined by
using trace norm distance in the set of states i.e. the quantity
$D_{1}(\ro)$ defined as \cite{Paula}
\begin{equation}
D_{1}(\ro)=\min\limits_{\PP^{\cA}}\,||\ro-\PP^{\cA}(\ro)||_{1},
\end{equation}
where $\PP^{\cA}$ is the projective measurement on subsystem $\cA$
i.e.
\begin{equation}
\PP^{\cA}(\ro)=\sum\limits_{k}(P_{k}\otimes \I)\,\ro\,(P_{k}\otimes
\I).
\end{equation}
Analytic expression for $D_{1}$ is known only for Bell - diagonal and $X$ - shaped
mixed states \cite{Cic}. In this note we study  $X$ - shaped two - qubit
states
\begin{equation}
\ro=\begin{pmatrix}\ro_{11}&0&0&\ro_{14}\\
0&\ro_{22}&\ro_{23}&0\\
0&\ro_{32}&\ro_{33}&0\\
\ro_{41}&0&0&\ro_{44}
\end{pmatrix},\label{X}
\end{equation}
where all matrix elements are real and non - negative. The quantity $D_{1}$ for such states can be computed
as follows. Let  $x=2(\ro_{11}+\ro_{22}) -1$ and
\begin{equation}
\al{1}=2(\ro_{23}+\ro_{14}),\quad \al{2}=2(\ro_{23}-\ro_{14}),\quad \al{3}=1-2(\ro_{22}+\ro_{33}).
\end{equation}
Then \cite{Cic}
\begin{equation}
D_{1}(\ro)=\sqrt{\frac{\DS a\,\al{1}^{2}-b\,\al{2}^{2}}{\DS a-b +\al{1}^{2}-\al{2}^{2}}},\label{disc1}
\end{equation}
where
\begin{equation}
a=\max\,(\al{3}^{2},\, \al{2}^{2}+x^{2}),\quad b=\min\,(\al{3}^{2},\, \al{1}^{2}).
\end{equation}
Notice that  the formula
(\ref{disc1}) is not valid in the case when $x=0$ and
\begin{equation}
|\al{1}|=|\al{2}|=|\al{3}|.
\end{equation}
In such a case, one can use general prescription how to compute
$D_{1}$, also given in Ref. \cite{Cic} (eq. (65)).
\par
Now we go to study the relations between $N(\ro),\, D_{1}(\ro)$ and
$\cm(\ro)$ for some specific classes of states of two qubits.
\subsection{Pure two qubit states}
Let us consider first the simple case of pure two qubit states. Up
to the local unitary operations, the corresponding density matrix
$\ro_{\mathrm{pure}}$ can be written as
\begin{equation}
\ro_{\mathrm{pure}}=\frac{1}{2}\,\begin{pmatrix}1+\sqrt{1-N^{2}}&0&0&N\\
0&0&0&0\\
0&0&0&0\\
N&0&0&1-\sqrt{1-N^{2}}\end{pmatrix},
\end{equation}
where $N\in [0,1]$. It is straightforward to show that in this case
\begin{equation}
N(\ro_{\mathrm{pure}})=D_{1}(\ro_{\mathrm{pure}})=N.
\end{equation}
Moreover the corresponding covariance matrix is of the form
\begin{equation}
Q=\begin{pmatrix}N&\hspace*{2mm}0&0\\0&-N&0\\0&\hspace*{2mm}0&N^{2}\end{pmatrix},
\end{equation}
so
\begin{equation}
\cm(\ro_{\mathrm{pure}})=||Q||=N,
\end{equation}
and
\begin{equation}
C(\ro_{\mathrm{pure}})=N+\frac{1}{2}N^{2}.
\end{equation}
Thus we arrive at the conclusion that for the correlated two qubit
pure states
\begin{equation}
N(\ro_{\mathrm{pure}})=D_{1}(\ro_{\mathrm{pure}})=\cm(\ro_{\mathrm{pure}})
< C(\ro_{\mathrm{pure}}).
\end{equation}
\subsection{Classically correlated mixed two qubit states}
We start the analysis of mixed two qubit states by considering so
called classical - quantum and classical - classical states. When we consider projective measurements
on subsystem $\cA$, there is a subclass of states which are left unperturbed by at least one such
measurement. Such "classical - quantum" states $\rocq$ have zero discord and are of the form
\begin{equation}
\rocq=\sum\limits_{k}p_{k}\, P_{k}^{\cA}\otimes \ro_{k}^{\cB},\label{rocq}
\end{equation}
where $\{P_{k}^{\cA}\}$ is a von Neumann measurement on $\cA$, $\{\ro_{k}^{\cB}\}$ are arbitray quantum states on $\cB$
and $\{p_{k}\}$ is the classical probability distribution. One can also consider fully classically correlated quantum states.
Such "classical - classical" states $\rocc$ can be written as
\begin{equation}
\rocc=\sum\limits_{j,k}p_{jk}P_{j}^{\cA}\otimes P_{k}^{\cB},\label{rocc}
\end{equation}
where $\{P_{j}^{\cA}\}$ and $\{P_{k}^{\cB}\}$ are von Neumann measurements on $\cA$ and $\cB$ respectively
and $\{p_{jk}\}$ is a two - dimensional probability distribution.  In the case of two qubits, these states are defined in terms
of orthogonal projectors
\begin{equation}
\begin{split}
&P_{1}=\begin{pmatrix} \cos^{2}\te&\frac{1}{2}e^{-i\vf}\,\sin 2\te\\[2mm]
\frac{1}{2}e^{i\vf}\,\sin 2\te&\sin^{2}\te\end{pmatrix},\\[2mm]
&P_{2}=\begin{pmatrix} \sin^{2}\te&-\frac{1}{2}e^{-i\vf}\,\sin 2\te\\[2mm]
-\frac{1}{2}e^{i\vf}\,\sin 2\te&\cos^{2}\te\end{pmatrix}.
\end{split}\label{P1P2}
\end{equation}
In particular
\begin{equation}
\rocq=p_{1}\,P_{1}\otimes \ro_{1}^{\cB}+p_{2}\,P_{2}\otimes
\ro_{2}^{\cB},\label{rocq2}
\end{equation}
where
\begin{equation}
\ro_{1,2}^{\cB}=\frac{1}{2}\left(\I+\tl{a}_{1,2}\cdot
\sib\right),\label{roB}
\end{equation}
and the vectors $\tl{a}_{1,2}\in \R^{3}$ satisfy $||\tl{a}_{1,2}||\leq 1$. Similarly, up to the local unitary
equivalence, the states $\ro_{cc}$ can be written as
\begin{equation}
\ro_{cc}=\sum\limits_{j,k=1}^{2}p_{jk}\,P_{j}\otimes
P_{k}.\label{rocc2}
\end{equation}
Since the states $\rocq$ and $\rocc$ have zero discord only
maximal mutual correlation and correlation distance can be greater then zero.
\par
Let us start with
classical - quantum state. By a direct calculations one shows that
\begin{equation}
\cm(\rocq)=2\,p_{1}p_{2}\, ||\tl{a}_{1}-\tl{a}_{2}||,\label{cmcq}
\end{equation}
where $||\cdot||$ is the euclidean norm in $\R^{3}$. The factor
$2\,p_{1}p_{2}$ in (\ref{cmcq}) can be interpreted in terms of the
statistical properties of a discrete classical random variable. Let
$X$ be the random variable with values in the set $\{-1,+1\}$ and
probability distribution
\begin{equation}
\mathrm{Prob}\{X=+1\}=p_{1},\quad \mathrm{Prob}\{X=-1\}=p_{2}.
\end{equation}
Then
\begin{equation}
\left(\mathrm{Var} X\right)^{2}=4\,p_{1}p_{2}
\end{equation}
and
\begin{equation}
\cm(\ro_{cq})=\frac{1}{2}\,\left(\mathrm{Var}
X\right)^{2}\,||\tl{a}_{1}-\tl{a}_{2}||.\label{cmrocq}
\end{equation}
Notice that in this case maximal mutual correlation depends on the classical probability distribution
defining the state $\rocq$ but also contains some information about the properties of quantum states of the
subsystem $\cB$. The largest possible value of $\cm(\rocq)$ is achieved for $p_{1}=p_{2}=1/2$ and when
the Bloch vectors
$\tl{a}_{1},\; \tl{a}_{2}$ are normalized and satisfy $\tl{a}_{2}=-\, \tl{a}_{1}$ i.e. when the states
 $\ro_{1}^{\cB},\; \ro_{2}^{\cB}$ are projectors on orthogonal one - dimensional subspaces. But such state
is the instance
 of classical - classical state (\ref{rocc2}), so for the genuine classical - quantum state $\cm(\rocq)<1$.
\par
In the case of general classical - classical state (\ref{rocc2}) the
calculations show that $\cm(\rocc)$ depends only on the properties
of classical probability distribution $\{p_{jk}\}$. One can check
that
\begin{equation}
\begin{split}
\cm(\rocc)=\big|&(p_{11}-p_{22})^{2}-(p_{12}-p_{21})^{2}\\
&+p_{12}+p_{21}-p_{11}-p_{22}\,\big|.\label{cmrocc}
\end{split}
\end{equation}
In this case, the right hand side of (\ref{cmrocc}) can be
interpreted as the modulus of covariance of two discrete random
variables $X$ and $Y$ with values in the set $\{-1,+1\}$. This can
be proved by considering the following join probability distribution
of $X$ and $Y$
\begin{equation}
\begin{split}
&\mathrm{Prob}\,\{X=+1,\, Y=+1\}=p_{11},\\
&\mathrm{Prob}\,\{X=+1,\, Y=-1\}=p_{12},\\
&\mathrm{Prob}\,\{X=-1,\, Y=+1\}=p_{21},\\
&\mathrm{Prob}\,\{X=-1,\, Y=-1\}=p_{11}.
\end{split}
\end{equation}
Since
\begin{equation}
\begin{split}
&\sr{X}=p_{11}+p_{12}-p_{21}-p_{22},\\
&\sr{Y}=p_{11}+p_{21}-p_{12}-p_{22}
\end{split}
\end{equation}
and
\begin{equation}
\sr{XY}=p_{11}+p_{22}-p_{12}-p_{21},
\end{equation}
then
\begin{equation}
\begin{split}
\mathrm{Cov}\,(X,Y)=&p_{11}+p_{22}-p_{12}-p_{21}\\
&+(p_{12}-p_{21})^{2}-(p_{11}-p_{22})^{2}.
\end{split}
\end{equation}
So
\begin{equation}
\cm(\rocc)=\big|\,\mathrm{Cov}\,(X,Y)\,\big|.
\end{equation}
Concerning the correlation distance, the states $\rocq$ and $\rocc$
have interesting property: in both cases correlation distance equals
to maximal mutual correlation
\begin{equation}
\cm(\rocq)=C(\rocq),\quad \cm(\rocc)=C(\rocc).
\end{equation}
In particular $C(\rocc)$ is given by covariance of classical random
variables.
\subsection{Some separable states with non - zero discord}
Consider now the following class of  two qubit states
\begin{equation}
\ro_{d}=\begin{pmatrix} w&0&0&s\\0&w&s&0\\0&s&\frac{1}{2}-w&0\\
s&0&0&\frac{1}{2}-w\end{pmatrix},\label{rod}
\end{equation}
where
\begin{equation}
0<w<\frac{1}{2}, \quad 0<s\leq s_{\mr{max}}
\end{equation}
and
\begin{equation}
s_{\mr{max}}=\sqrt{\frac{1}{2}w-w^{2}}.
\end{equation}
The states (\ref{rod}) are separable, but
\begin{equation}
D_{1}(\ro_{d})=\frac{\DS 4s\,|1-4w|}{\DS \sqrt{16s^{2}+(1-4w)^{2}}}.
\end{equation}
Thus for all $w\in (0,1/2)$ except of $w=1/4$ and admissible $s$,
$\ro_{d}$ has non - zero discord. On the other hand, for such states
the covariance matrix $Q$ is very simple and equals to
\begin{equation}
Q=\begin{pmatrix}4s&0&0\\0&0&0\\0&0&0\end{pmatrix},
\end{equation}
so
\begin{equation}
\cm(\ro_{d})=C(\ro_{d})=4s
\end{equation}
and
\begin{equation}
D_{1}(\ro_{d})<\cm(\ro_{d})=C(\ro_{d})
\end{equation}
\subsection{Some entangled states }
Now we take the states which are entangled. Consider the following family of
two qubit states \cite{Mizra}
\begin{equation}
\ro_{\te}=\begin{pmatrix}\frac{1}{2}\cos^{2}\te&0&0&\frac{1}{4}\sin 2\te\\[2mm]
0&0&0&0\\[2mm]
0&0&\frac{1}{2}&0\\[2mm]
\frac{1}{4}\sin 2\te&0&0&\frac{1}{2}\sin^{2}\te
\end{pmatrix},\label{rote}
\end{equation}
where $\te \in (0,\pi/2)$. For this family we have
\begin{equation}
N(\ro_{\te})=\frac{\DS \sqrt{6-2\cos 4\te}-2}{\DS 4}
\end{equation}
and
\begin{equation}
D_{1}(\ro_{\te})=\frac{1}{2}\sin 2\te.
\end{equation}
Moreover, the covariance matrix equals to
\begin{equation}
Q=\begin{pmatrix} \frac{1}{2}\sin 2\te&0&0\\[2mm]0&-\frac{1}{2}\sin 2\te&0\\[2mm] 0&0&\frac{1}{4}\sin^{2} 2\te
\end{pmatrix},
\end{equation}
so
\begin{equation}
\cm(\ro_{\te})=\frac{1}{2}\sin 2\te
\end{equation}
and
\begin{equation}
C(\ro_{\te})=\frac{1}{2}\sin 2\te +\frac{1}{8}\sin^{2} 2\te.
\end{equation}
Thus
\begin{equation}
N(\ro_{\te})<D_{1}(\ro_{\te})=\cm(\ro_{\te})<C(\ro_{\te})
\end{equation}
for all $\te\in (0,\pi/2)$.
\subsection{General class of Bell - diagonal states}
Consider now the states of the form
\begin{equation}
\ro_{\mathrm{BD}}=\frac{1}{4}\,\left(\I\otimes \I+\sum\limits_{j=1}^{3}c_{j}\,\si{j}\otimes\si{j}\,\right),\label{BD}
\end{equation}
where $\tl{c}=(c_{1},\, c_{2},\, c_{3})$ is a three - dimensional real vector. Define also
\begin{equation}
c_{+}=\max\, \left(|c_{1}|,\, |c_{2}|,\, |c_{3}|\right)
\end{equation}
and
\begin{equation}
c_{0}=\mathrm{int}\,\left(|c_{1}|,\, |c_{2}|,\, |c_{3}|\right)\label{int}
\end{equation}
where (\ref{int}) represents intermediate among the numbers $|c_{1}|,\, |c_{2}|$ and $|c_{3}|$. It is known  that \cite{Paula}
\begin{equation}
D_{1}(\ro_{\mathrm{BD}})=c_{0}
\end{equation}
and
\begin{equation}
N(\ro_{\mathrm{BD}})\leq D_{1}(\ro_{\mathrm{BD}}).
\end{equation}
On the other hand, corresponding correlation matrix $Q$ is also diagonal and reads
\begin{equation}
Q=\begin{pmatrix} c_{1}&0&0\\0&c_{2}&0\\0&0&c_{3}\end{pmatrix},
\end{equation}
so
\begin{equation}
\cm(\ro_{\mathrm{BD}})=c_{+}.
\end{equation}
Since $c_{0}\leq c_{+}$, we obtain
\begin{equation}
N(\ro_{\mathrm{BD}})\leq D_{1}(\ro_{\mathrm{BD}})\leq \cm(\ro_{\mathrm{BD}})\leq C(\ro_{\mathrm{BD}}).
\end{equation}
\section{Conclusions}
In this note we have studied bipartite correlations in the algebraic
formulation of quantum mechanics. The proposed  measure of
correlations, called maximal mutual correlation, is a generalization
of a natural measure of nonseparability of pure states. In the case
of pure qubit states, such measure equals to entanglement, but for
mixed states it can be non - zero also for classically correlated
quantum states. As we have shown on explicit examples, maximal
mutual correlation is  greater or equal to quantum discord and is
always dominated by the correlation distance.

\end{document}